\documentclass[aps,prl,reprint,groupedaddress]{revtex4-1}
\usepackage{graphicx} 
\usepackage{dcolumn} 
\usepackage{bm} 
\usepackage{amssymb} 
\usepackage{draftcopy}
\usepackage{subfigure}
\usepackage{epstopdf}
\usepackage{units}
\begin{document}
\newcommand{\EuYSO}{Eu$^{3+}$:Y$_2$SiO$_5$~}
\newcommand{\Eu}{Eu$^{3+}$~}
\newcommand{\YSO}{Y$_2$SiO$_5$}
\newcommand{\SHBstab}{$1.0\times10^{-15}~ \tau^{\nicefrac{-1}{2}}$~}
\newcommand{\SHBabs}{$8.5^{+4.8}_{-1.8}\times10^{-17}$ at $\tau = 73$ s}
\newcommand{\SHBxvx}{$4.6^{+1.4}_{-0.7}\times10^{-17}$ at $\tau = 2000$~}
\newcommand{\SHBram}{$1\times10^{-16}$~}
\newcommand{\SHBramabs}{$1\times10^{-6}$~}
\title{Laser frequency stabilization based on steady-state spectral-hole burning in \EuYSO}
\author{Shon~Cook}
\email{shon.cook@nist.gov}
\author{Till~Rosenband}
\author{David~R.~Leibrandt}
\affiliation{National Institute of Standards and Technology, 325 Broadway St., Boulder, CO 80305, USA}
\date{\today}
\begin{abstract}
We present and analyze a method of laser frequency stabilization via steady-state patterns of spectral holes in Eu$^{3+}$:Y$_2$SiO$_5$. Three regions of spectral holes are created, spaced in frequency by the ground state hyperfine splittings of $^{151}$Eu$^{3+}$. The absorption pattern is shown not to degrade after days of laser frequency stabilization.  An optical frequency comparison of a laser locked to such a steady-state spectral-hole pattern with an independent cavity-stabilized laser and a Yb optical lattice clock demonstrates a spectral-hole fractional frequency instability of \SHBstab that averages to \SHBabs. Residual amplitude modulation at the frequency of the RF drive applied to the fiber-coupled electro-optic modulator is reduced to less than \SHBramabs  fractional amplitude modulation at $\tau>$ 1 s by an active servo.  The contribution of residual amplitude modulation to the laser frequency instability is further reduced by digital division of the transmission and incident photodetector signals to less than \SHBram at $\tau>$ 1 s.
\end{abstract}
\pacs{42.62.Eh, 42.62.Fi, 78.47.nd​}
\keywords{laser frequency stabilization, spectral hole burning, spectral diffusion}
\maketitle
Frequency stable laser local oscillators (LLOs) are key tools in the field of metrology. Applications of such LLOs include optical atomic clocks \cite{bloom2014optical, CWC2010AlAl, jiang2011making}, tests of general relativity \cite{abbott2009ligo, reynaud2009testing}, searches for variation of fundamental constants \cite{Blatt2008SrGravity, TR2008AlHg}, and relativistic geodesy \cite{chou2010optical, bondarescu2012geophysical}. The best stabilized lasers to date \cite{kessler2012sub, hafner20158e} are obtained by locking their frequencies to Fabry-P\'{e}rot reference cavities, and their stability is intrinscially limited by thermomechanical length fluctuations of the cavity \cite{Numata2004ThermalNoise, kessler2012thermal}.

Higher stability might be possible by the method of spectral-hole burning (SHB) \cite{sellin2001laser, pryde2002semiconductor, Strickland2000}, where a narrow optical transition of rare-earth ions doped into a cryogenically cooled crystal is used as the frequency reference. Strains in the crystalline host inhomogenously broaden the narrow absorption line by several gigahertz. For laser frequency stabilization, narrow transparencies (spectral holes) are created within the absorption line by frequency selective optical pumping \cite{julsgaard2007understanding}. These narrow spectral holes can be used for laser frequency stabilization by actively steering the frequency of the laser to maximize the transmission through the crystal.

Spectral holes in \EuYSO \cite{konz2003temperature} are particularly promising for laser frequency stabilization. At 4 K, this material supports spectral holes at 580 nm with linewidths as narrow as 122 Hz \cite{equall1994ultraslow} and lifetimes of $10^6$ s \cite{konz2003temperature}. The frequency shifts due to fluctuations in the ambient magnetic and electric fields, temperature, pressure, and acceleration are all small enough to allow laser frequency instabilities at the $10^{-17}$ fractional frequency level \cite{thorpe2011frequency, thorpe2013shifts}.  However, prior laser frequency stabilization experiments with \EuYSO have been limited to run-times of a few thousand seconds due to degradation of the spectral holes caused by noise and saturation of the probe laser \cite{leibrandt2013absolute}.

\begin{figure}
\includegraphics[angle=270,width=1.0\columnwidth]{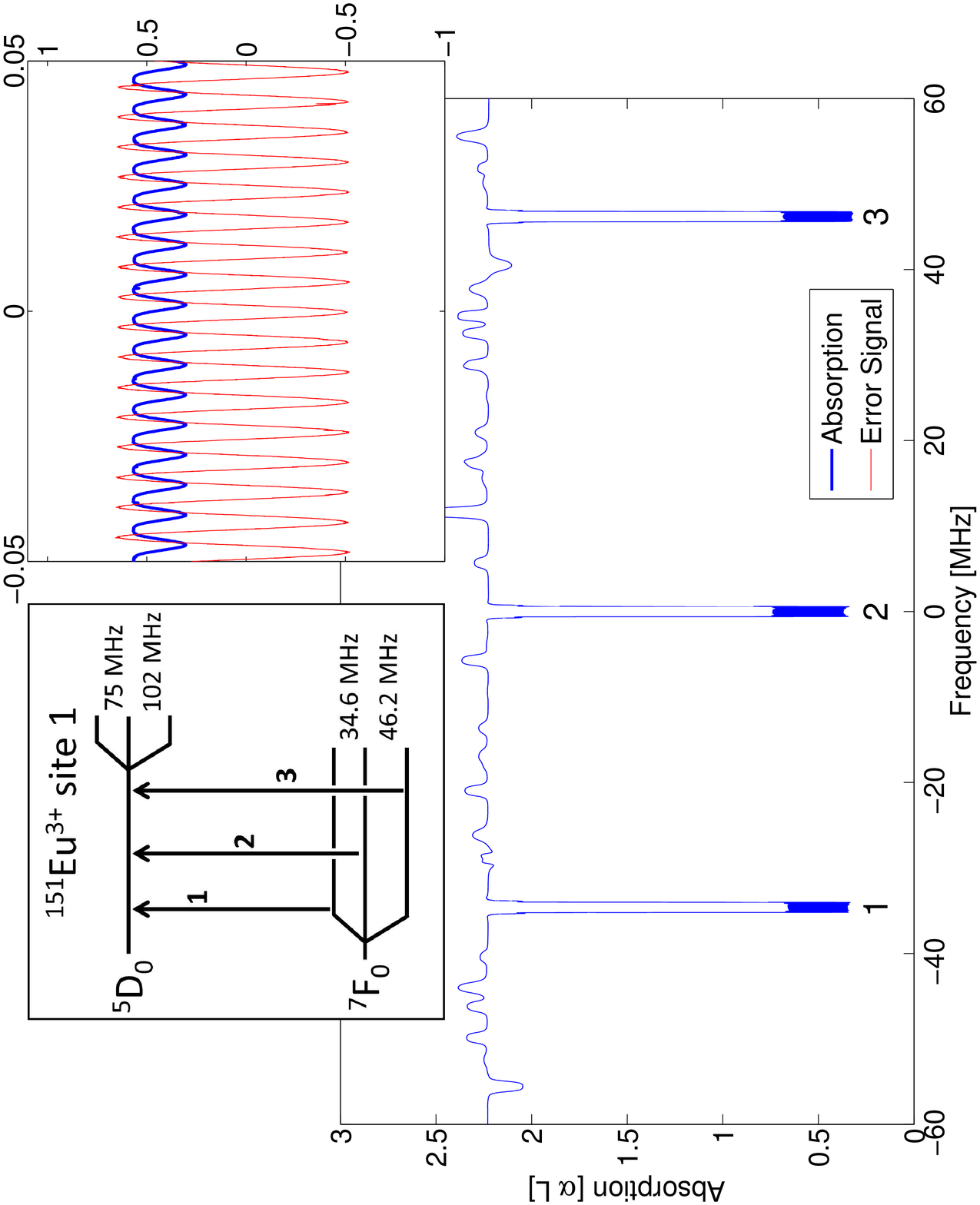}
\caption{\label{fig3} Absorption of a steady-state spectral hole pattern with three regions of spectral holes.  The left inset shows a level diagram of $^{151}$Eu$^{3+}$ positioned at crystallographic site 1.  The right inset shows the absorption (blue) and corresponding PDH error signal (red) of a subset of the center region.}
\end{figure}

In this work, we demonstrate frequency stabilization to a steady-state pattern of spectral holes in 1.0 atomic~\% \EuYSO crystals at temperatures near 4~K.  This pattern consists of three regions of spectral holes spaced in frequency by 46.2 MHz and 34.6 MHz (see Figure~\ref{fig3}), corresponding to the ground-state hyperfine splittings of $^{151}$Eu$^{3+}$ positioned at crystallographic site 1 \cite{yano1991ultralong,yano1992hyperfine}.  The \Eu population reaches steady-state as the spectral holes in the three regions are burned.  Additional interleaved probing does not modify the absorption spectrum because, for a subset of the $^{151}$Eu$^{3+}$ ions with certain detunings within the inhomogeneously broadened absorption line, all three of the hyperfine ground states are addressed by the laser as it probes spectral holes in the three regions.  Absorption does not accumulate between the spectral holes because they are spaced by only a few times the minimum observed spectral-hole width.  This persistent and self-regenerating  pattern is achieved by utilizing probe pulses whose center frequency coincides with an absorption minimum.  Weaker phase modulation sidebands are spaced to coincide with  absorption maxima to yield a Pound-Drever-Hall type error signal.  Using this spectral-hole pattern, laser frequency stabilization experiments can be made to run indefinitely.

\begin{figure}
\includegraphics[width=1.0\columnwidth]{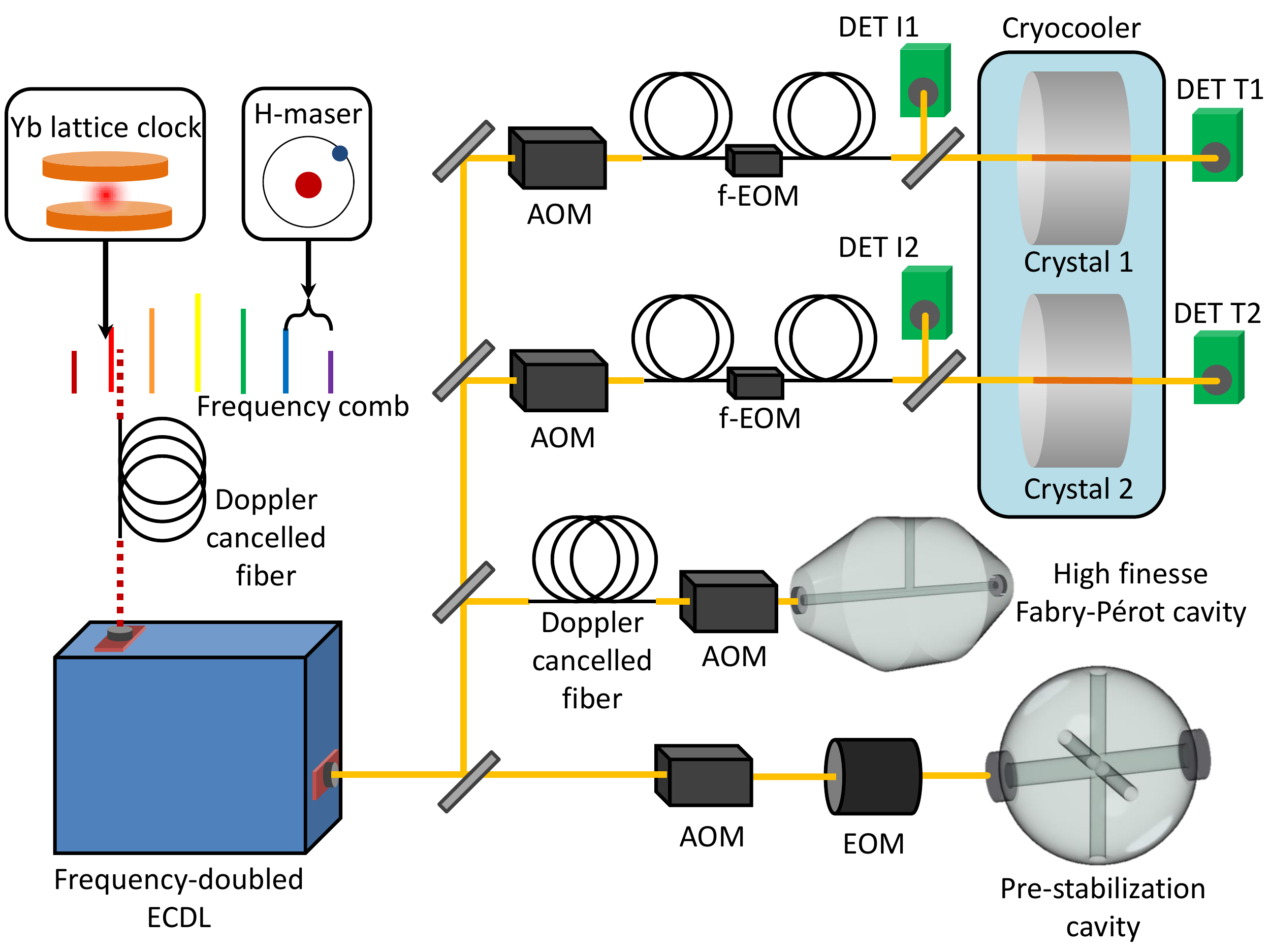}
\caption{\label{fig1} Schematic diagram of the spectral-hole burning experiment.  Solid yellow lines denote 580 nm laser propagation, and dashed red lines denote 1160 nm laser propagation. AOM: acousto-optic modulator, f-EOM: fiber-coupled electro-optic modulator, DET: detector, ECDL: external cavity diode laser.}
\end{figure}

\begin{figure}
\includegraphics[width=1.0\columnwidth]{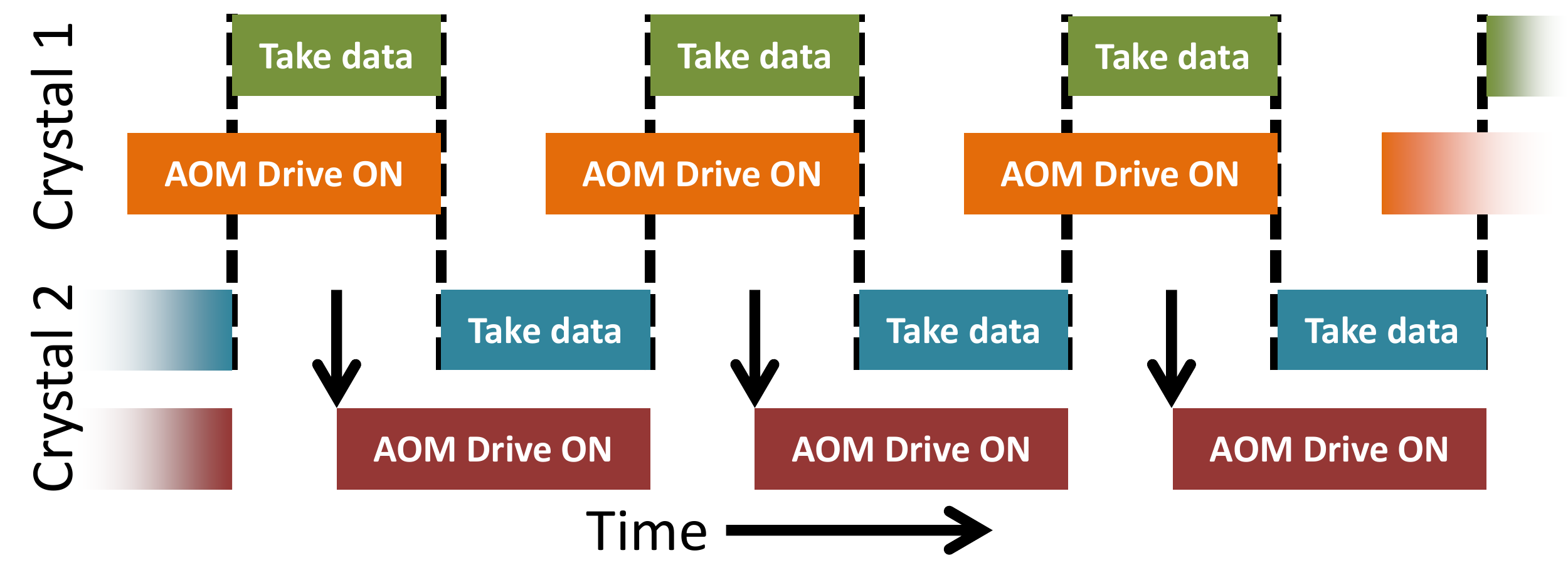}
\caption{\label{fig1c} Pulse sequence of the interleaved probe scheme for 100\% duty cycle laser frequency stabilization.  Frequency feedback from the previous probe pulse on each crystal is applied to the pre-stabilization cavity AOM at the times indicated by vertical arrows.}
\end{figure}

\begin{figure}
\includegraphics[width=1.0\columnwidth]{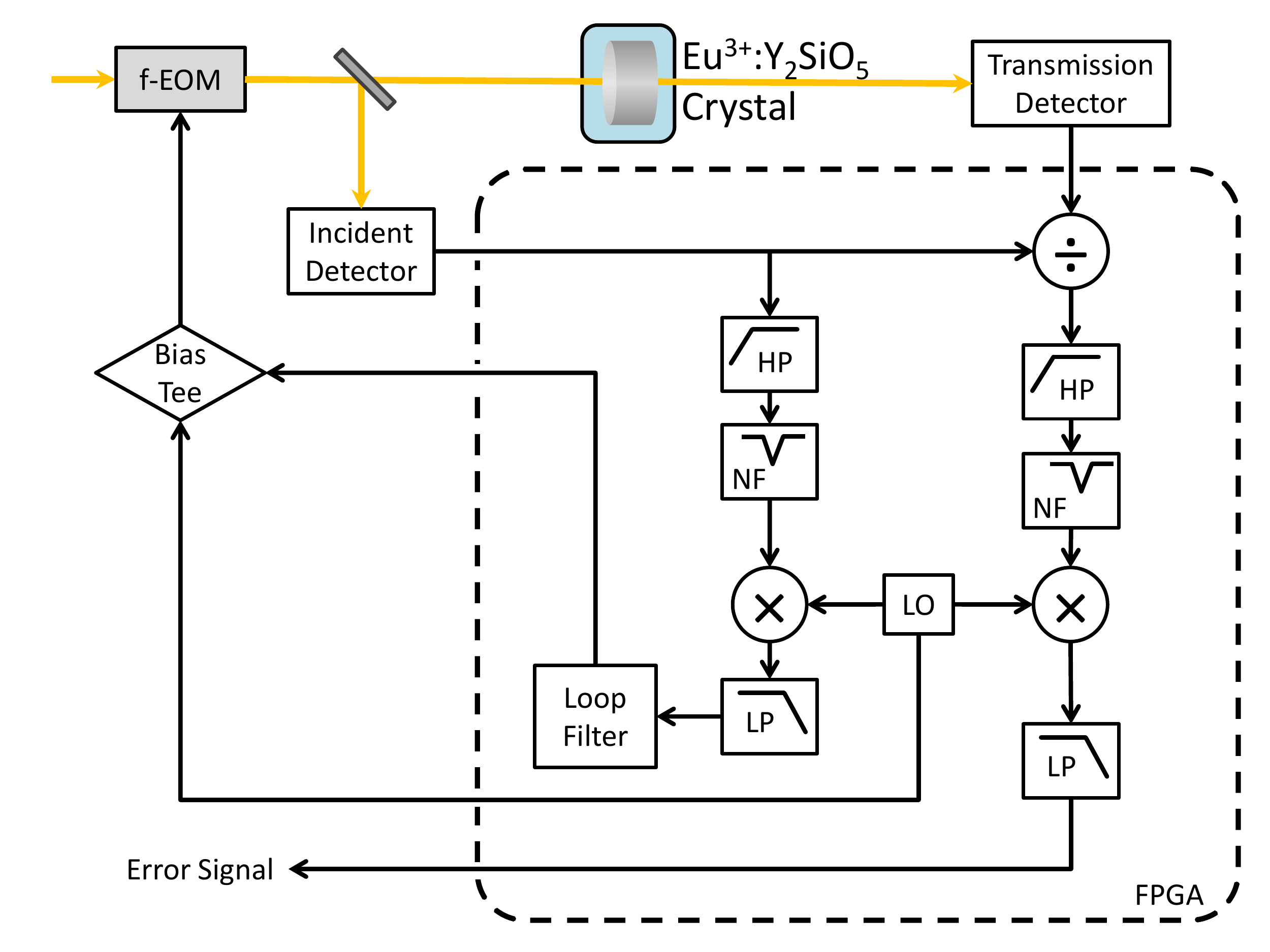}
\caption{\label{fig1b} Frequency feedback and residual amplitude modulation servo schematic diagram.  Black arrows denote electronic signal propagation, and the dashed box denotes calculations performed digitally on a field-programmable gate array (FPGA). LP: low-pass filter, HP: high-pass filter, LO: local oscillator, NF: notch filter.}
\end{figure}


Our SHB LLO employs a two-stage laser frequency stabilization scheme, as shown in Figure \ref{fig1} and similar to that described in Reference \cite{leibrandt2013absolute}. A 1160 nm external cavity diode laser (ECDL) is frequency doubled to 580 nm, and pre-stabilized to a Fabry-P\'{e}rot cavity \cite{leibrandt2013cavity}.  The pre-stabilized laser fractional frequency instability is $3\times10^{-15}$ for 0.1 s $< \tau <$ 1 s.  This light is split into two beams that are sent through two independent frequency and intensity tuneable acousto-optic modulators (AOMs) in order to burn and probe spectral holes in two \EuYSO crystals. A laser intensity of 2 $\mu$w/cm$^2$ incident on the crystals is used for both burning and probing for all of the experiments discussed in this letter.  Two fiber-coupled electro-optic modulators (f-EOMs) apply sidebands to the beams at 15~kHz for crystal one and 21~kHz for crystal two, for generation of a Pound-Drever-Hall (PDH) error signal \cite{Drever1983PDH, julsgaard2007understanding}. Detectors (DETs) I1 and I2 measure the power incident on the two crystals, which is stabilized by feedback to the RF power of the probe AOM drives. Detectors T1 and T2 measure the probe light transmitted through the crystals.  For measurements of the laser frequency stability, visible laser light is sent to an independent high-finesse Fabry-P\'{e}rot reference cavity \cite{BCY1999subhertz}, and infrared laser light is sent to a frequency comb for comparison against other frequency references, each through a Doppler-cancelled optical fiber \cite{Ma1994FiberNoise}.

During a SHB laser frequency stabilization run, several hundred spectral holes are created. A hole is then randomly selected, and the laser spends a few milliseconds probing the center of the spectral hole.  Frequency corrections from the two crystals are averaged and applied to the pre-stabilization cavity AOM.  In order to reduce laser frequency degradation due to the Dick effect \cite{dick1990local}, probe pulses for the two crystals are interleaved with a 100\% duty cycle, as shown in Figure \ref{fig1c}.  To reduce frequency instability caused by error signal transients at the beginning of each probe pulse, the error signal generated during the first several hundred microseconds of each pulse is not used for frequency feedback.

A schematic diagram of the error signal generation and residual amplitude modulation (RAM) servo is shown in Figure \ref{fig1b}. The transmission detector signal is digitally divided by the incident detector signal on a field-programmable gate array (FPGA) in order to suppress laser intensity noise.  This also provides some suppression of the RAM caused by the f-EOM. After filtering, this signal is multiplied with the local oscillator to produce the PDH error signal. The incident detector signal is independently demodulated in order to control the RAM through a servo-loop by applying a DC bias to the f-EOM \cite{zhang2014reduction}.  This suppresses the contribution of RAM, converted to fractional frequency instability of a laser locked to a steady-state hole pattern, to less than $1\times10^{-16}$ at averaging times greater than 1~s, and to less than $1\times10^{-17}$ at averaging times greater than 100~s.

To reduce frequency instability due to temperature fluctuations \cite{thorpe2013shifts}, the \EuYSO crystals are housed in a sealed copper pillbox with a fixed amount of helium gas.  The temperature of the cryocooler is servo controlled to keep the copper pillbox within 5 mK of the first-order insensitive temperature (3.941~K).  The measured out-of-loop temperature instability, converted to fractional frequency instability, is less than $1\times10^{-16}$ at averaging times greater than 1~s, and less than $1\times10^{-17}$ at averaging times between 100~s and $10^4$~s.

\begin{figure}
\includegraphics[angle=270,width=1.0\columnwidth]{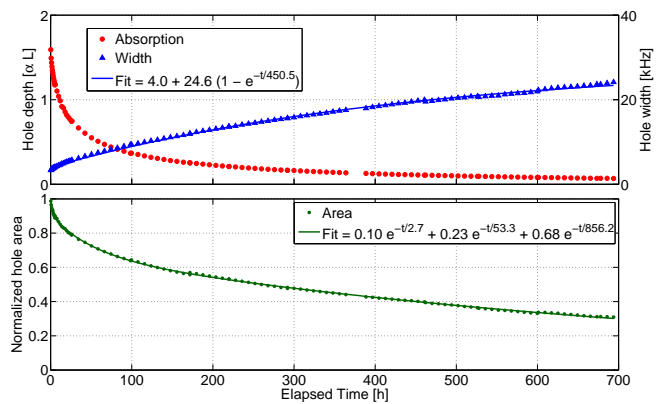}
\caption{\label{fig6} Time evolution of a spectral hole in the absence of degradation due to probing.  The top plot shows a decrease in the absorption depth (red circles, left axis) and an increase in the full-width at half-maximum (FWHM) (blue triangles, right axis) with a fit of the width to a function $A + B(1-e^{-t/\tau})$.  The bottom plot shows the area of the spectral hole normalized to the initial area.  The data is fit to a three-part exponential decay.}
\end{figure}

The time-evolution of an unperturbed spectral hole at 3.941 K is shown in Figure \ref{fig6}.  The top plot of Figure \ref{fig6} shows a decrease in hole depth (in units of $\alpha$L from maximum absorption of the crystal) and an increase in hole width.  The bottom plot of Figure \ref{fig6} shows the time-evolution of the normalized area of the spectral-hole.  The hole area is fit to a three-part exponential decay with empirically determined time constants of 2.7 h, 53.3 h, and 856.2 h.  During a steady-state laser frequency stabilization run, the entire spectral-hole pattern must be probed frequently enough to re-burn all of the spectral holes before they decay.  Spectral diffusion \cite{bai1989time, yano1992stimulated} causes the spectral holes to broaden, limiting the minimum frequency spacing of holes within the steady-state pattern.

\begin{figure}
\includegraphics[angle =270,width=1.0\columnwidth]{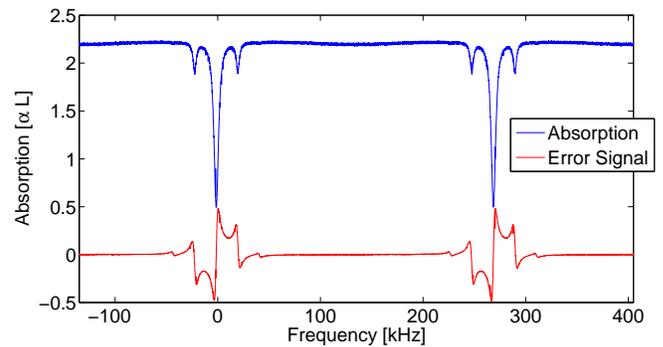}
\caption{\label{fig2} Absorption (blue) and corresponding PDH error signal (red) of two out of a total of 603 spectral holes in a single-region spectral-hole pattern.}
\end{figure}

The absorption pattern of the three-region steady-state spectral-hole pattern after three hours of laser frequency stabilization run-time is shown in Figure \ref{fig3}.  Each of the three regions has 201 holes spaced by 6 kHz.  As the pattern reaches steady-state, the spectral holes increase from an initial 1.7 kHz full-width at half-maximum (FWHM) to their steady-state FWHM of 3.3 kHz after ten hours of run-time.

In contrast, a subset of the single-region spectral-hole pattern used in previous work \cite{leibrandt2013absolute} is shown in Figure \ref{fig2}, after three hours of run-time.  The 603 holes of this pattern are spaced by 270 kHz.  Starting from an initial width similar to that of the three-region pattern, the spectral holes have broadened to 5.6~kHz after three hours of run-time.  The spectral-holes continue to broaden indefinitely, which degrades the frequency stability of the laser locked to this single-region spectral-hole pattern.

\begin{figure}
\includegraphics[angle=270,width=1.0\columnwidth]{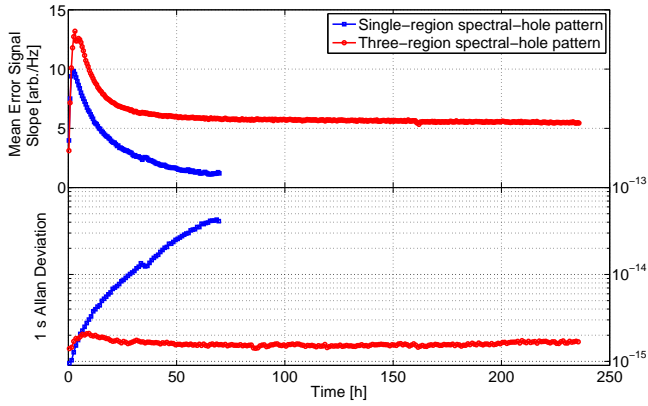}
\caption{\label{fig4} Comparison of a single-region pattern (blue squares) versus a three-region steady-state spectral-hole pattern (red circles).  The top plot shows the error signal slope averaged over all of the holes in the pattern, and the bottom is the 1 second crystal versus crystal Allan deviation with linear frequency drift removed.  Both are plotted with respect to the run-time of the laser frequency stabilization.}
\end{figure}

Interspersed periodically between probe pulses used for laser frequency stabilization are scan pulses of individual spectral holes, which allow for the hole shape and error signal slope to be monitored as a run progresses.  The top plot of Figure \ref{fig4} shows a comparison between the slope of the error signal of a single-region pattern and a three-region steady-state pattern, as a function of the duration of time that the pattern has been used for laser frequency stabilization.  These values are an average of the error signal slope at the center of each spectral hole within the pattern.  For both patterns, the error signal slope reaches a maximum as the absorption at the center of the holes approaches zero, then decreases as the holes start to broaden.  The three-region steady-state pattern asymptotically approaches a non-zero error signal slope, with a time constant that indicates how long it takes the pattern to reach steady-state.  The single-region pattern error signal slope decreases until the lock can no longer be maintained after approximately 70 hours of run-time.  The bottom plot of Figure \ref{fig4} shows the 1 second averaging time Allan deviation of the differential frequency noise of the two laser beamlines independently locked to the two crystals, after linear drift has been removed.  The Allan deviation values have been divided by $\sqrt{2}$, which is an upper bound on the cystal with better spectral-hole frequency stability.  After the three-region pattern has reached steady-state, the 1 second differential fractional frequency instability never rises above $2\times10^{-15}$, out to a run-time of almost ten days.  The differential frequency instability of the single-region pattern steadily increases, rising above $1\times10^{-14}$ at 33 hours of run-time.

\begin{figure}
\includegraphics[angle=270,width=1.0\columnwidth]{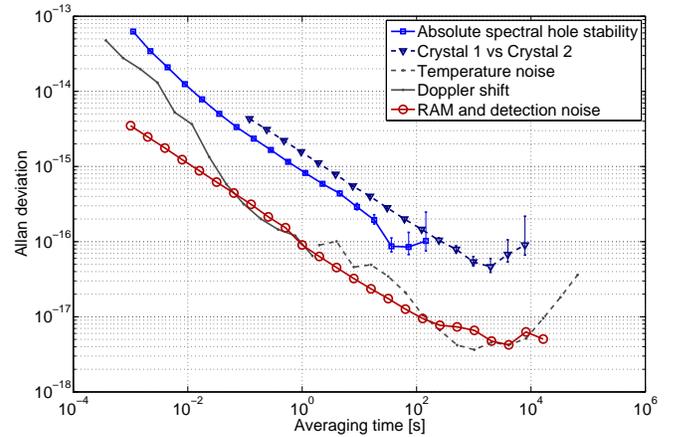}
\caption{\label{fig5} Fractional frequency instability of a laser locked to a steady-state pattern of spectral holes in \EuYSO.  Absolute laser frequency instability from a three-cornered hat measurement is shown (blue squares), as well as the differential instability of a crystal versus crystal frequency comparison (triangles) where much of the technical noise is common-mode.  The instability of the crystal versus crystal measurement is worse at small averaging times because the duty-cycle is 50\%, while the absolute spectral hole stability measurement uses feedback from both crystals to achieve 100\% duty cycle.  The expected frequency noise due to temperature fluctuations (dashed grey), Doppler shifts due to motion of the pillbox (solid grey), as well as RAM and detection noise (open red circles) are also shown.}
\end{figure}

We have measured the frequency stability of a laser locked to a three-region steady-state pattern of spectral holes in \EuYSO using a three-cornered hat method \cite{gray1974method}.  The stability is measured relative to an independent Fabry-P\'{e}rot reference cavity \cite{BCY1999subhertz} and a Yb optical lattice clock with a stability of $3.2\times10^{-16}  ~\tau^{\nicefrac{-1}{2}}$ \cite{hinkley2013atomic}.  The fractional frequency instability of the laser stabilized to the spectral holes is \SHBstab for 0.01 s $ < \tau < $ 20 s, and reaches \SHBabs, as shown in Figure~\ref{fig5}.  The duration of this measurement was limited by the duration of time the frequency comb remained mode-locked.  The crystal versus crystal fractional frequency instability averages to \SHBxvx s with a 0.5 mHz/s linear drift removed.  This Allan deviation of the frequency difference between the two crystals has been divided by $\sqrt{2}$ to put an upper bound on the frequency stability of the better of the two crystals.  Figure \ref{fig5} also shows the estimated contribution of temperature instability, Doppler shifts, RAM, and detection noise to the frequency instability of the SHB LLO.  Temperature instability is measured using an out-of-loop sensor and converted to frequency instability assuming operation at 5 mK away from the first-order insensitive temperature.  Doppler shifts introduced by the motion of the cryostat are measured interferometrically by retroreflecting a laser off of a mirror mounted on the copper pillbox.  These sources of laser frequency instability are shown not to limit the present performance of this spectral-hole burning experiment.

We have demonstrated steady-state SHB laser frequency stabilization with a short-term fractional frequency instability of \SHBstab that averages down to \SHBabs. The steady-state spectral-hole pattern successfully prevents degradation of the laser frequency stability out to run-times of many days. RAM is suppressed to less than \SHBramabs fractionally at 1~s averaging time using an active servo, and the digital division of transmitted and incident photodetector signals further suppresses laser frequency instability due to RAM.  

We thank J.~Bergquist, R. ~Fox, and D.~Wineland for useful discussions. Special thanks go to T. Fortier and S. Diddams for operation of the frequency comb used in the three-cornered hat measurement, and to the NIST Yb optical lattice clock team.  This work is supported by the Defense Advanced Research Projects Agency and the Office of Naval Research, and is not subject to US copyright.

\end{document}